\pdfoutput=1
\documentclass[%
reprint,
superscriptaddress,
bibnotes,
amsmath,amssymb,
pra,
floatfix,
]{revtex4-1}
\usepackage{graphicx}
\usepackage{dcolumn}
\usepackage{bm}
\usepackage{hyperref}

\usepackage{natbib}
\usepackage{mathrsfs,mathtools,color,wasysym,dsfont,here}
\usepackage{amsthm}
\usepackage{physics}
\usepackage[all]{xy}
\usepackage{tikz}
\usepackage{amsfonts}
\usepackage{array,booktabs}
\usepackage{comment}
\usetikzlibrary{calc}
\usetikzlibrary{arrows}
\usetikzlibrary{decorations.markings,decorations.pathmorphing}
\usepackage{cases}

\AtBeginDocument{
\heavyrulewidth=.08em
\lightrulewidth=.05em
\cmidrulewidth=.03em
\belowrulesep=.65ex
\belowbottomsep=0pt
\aboverulesep=.4ex
\abovetopsep=0pt
\cmidrulesep=\doublerulesep
\cmidrulekern=.5em
\defaultaddspace=.5em
}

\preprint{}

\newcommand{\waveq}{\ensuremath{q}}

\begin{document}

\title{Liouvillian gap and single spin-flip dynamics in the dissipative Fermi-Hubbard model}

\author{Hironobu Yoshida}
\email{hironobu-yoshida57@g.ecc.u-tokyo.ac.jp}
\affiliation{Department of Physics, Graduate School of Science, The University of Tokyo, 7-3-1 Hongo, Tokyo 113-0033, Japan}

\author{Hosho Katsura}
\affiliation{Department of Physics, Graduate School of Science, The University of Tokyo, 7-3-1 Hongo, Tokyo 113-0033, Japan}
\affiliation{Institute for Physics of Intelligence, The University of Tokyo, 7-3-1 Hongo, Tokyo 113-0033, Japan}
\affiliation{Trans-scale Quantum Science Institute, The University of Tokyo, 7-3-1, Hongo, Tokyo 113-0033, Japan}
\begin{abstract}
Motivated by recent progress in cold-atom experiments, we analyze the SU($N$) Fermi-Hubbard model on a $d$-dimensional hypercubic lattice with two-body loss. By focusing on states near the ferromagnetic steady states, we obtain the Liouvillian gap in closed form for any $d$ and $N$. 
We also investigate the dynamics of a ferromagnetic initial state with a single spin flip both analytically and numerically. 
In particular, we show that, by decreasing the strength of the interaction and loss, the survival probability of the spin flip exhibits a crossover from the power-law decay to the exponential decay. We expect that our findings can be tested experimentally with ultracold alkaline-earth-like atoms in an optical lattice.
\end{abstract}

\maketitle

\section{Introduction}
With recent advances in quantum engineering, it becomes increasingly important to investigate the effect of dissipation on the dynamics of quantum many-body systems. Ultracold atom experiments provide an ideal playground to address this issue, because the coupling to the environment and the particle loss can be well-controlled. For example, the Fermi-Hubbard model with SU($N$) spin symmetry is realized with alkaline-earth-like atoms~\cite{cazalilla_ultracold_2009,gorshkov_two-orbital_2010,Taie2010,Taie2012,Desalvo2010,Lewenstein2012,Scazza2014,Zhang2014,Cazalilla2014,Pagano2014,Hofrichter2016}, and well controlled two-body loss can be induced by photoassociation
~\cite{yan_observation_2013,zhu_suppressing_2014,sponselee_dynamics_2018,honda_observation_2023}. Motivated by these experiments, the dynamics of the Fermi-Hubbard model with two-body loss has been studied theoretically~\cite{sponselee_dynamics_2018,booker_non-stationarity_2020,rosso_eightfold_2022,mazza_dissipative_2022}.

Under the Markov approximation, the dynamics of an open quantum system is generated by the Liouvillian superoperator, and thus the inverse of the gap of the Liouvillian spectrum can be identified with the asymptotic decay rate, though some exceptions are known~\cite{mori_resolving_2020,haga_liouvillian_2021}. In one dimension, the Liouvillian gap of the SU($2$) Fermi-Hubbard model has been calculated exactly~\cite{nakagawa_exact_2021}, and the dynamics of the model was investigated with the loss rate equation~\cite{sponselee_dynamics_2018,rosso_eightfold_2022,rosso_dynamical_2023} and the quantum trajectory method~\cite{foss-feig_steady-state_2012,nakagawa_dynamical_2020}. However, the analysis in higher dimensions is notoriously difficult, because analytical tools such as Bethe ansatz are inapplicable and numerical approaches are computationally costly. 

In this paper, we study the SU($N$) Fermi-Hubbard model on a $d$-dimensional hypercubic lattice with two-body loss. In particular, we focus on states near the steady state. 
At $1/N$ filling (one particle per site), 
these states are written as a superposition of spin-flip excitations and doublon-hole excitations, allowing us to obtain the Liouvillian gap in closed form for any $d$ and $N$. As a result, we find that the Liouvillian gap of the system with linear size $L$ is proportional to $1/L^2$ but does not depend on $d$ or $N$. 

We also elucidate the dynamics 
of a ferromagnetic initial state with a single spin flip using the same method and obtain 
the analytical expression for
the survival probability of the spin flip in two limits: (i) strongly interacting and dissipative limit and (ii) weakly interacting and dissipative limit. In limit (i), it shows the power-law decay with a normalized timescale that is relevant to the continuous quantum Zeno effect~\cite{syassen_strong_2008,zhu_suppressing_2014,daley_quantum_2014,nakagawa_dynamical_2020, rosso_dynamical_2023}. On the other hand, in limit (ii), it shows the exponential decay. We also numerically checked that by decreasing the strength of the interaction and loss, a crossover from the power-law decay to the exponential decay occurs. 


\section{The model and methods}
\subsection{The model}
We consider $N$-component fermions on a $d$-dimensional hypercubic lattice $\Lambda$ with linear size $L$ and 
volume $L^d$ with periodic boundary conditions. For each site $\bm{x}=(x_1,\ldots,x_d)\in \Lambda$ $(x_j=0,\ldots, L-1)$, we denote by $\hat{c}^\dagger_{\bm{x},\sigma}$ and $\hat{c}_{\bm{x},\sigma}$ the creation and annihilation operators, respectively, of a fermion with spin $\sigma=1, \ldots, N$. The number operators are defined as $\hat{n}_{\bm{x},\sigma} :=\hat{c}^\dagger_{\bm{x},\sigma} \hat{c}_{\bm{x},\sigma}$ and $\hat{N}_f:=\sum_{\bm{x}\in \Lambda} \sum_{\sigma=1}^{N} \hat{n}_{\bm{x},\sigma}$. We use $N_f$ and $n_f$, respectively, to denote 
the total fermion number and the fermion density, i.e., $n_f=N_f/L^d$.
Under the Markov approximation, the dynamics of the system with two-body loss is described by the Gorini-Kossakowski-Sudarshan-Lindblad (GKSL) master equation~\cite{lindblad_generators_1976, gorini_completely_1976,breuer_theory_2007}
\begin{equation}
    \frac{d \hat{\rho}}{d t}=\mathcal{L}\hat{\rho}:=-i[\hat{H}, \hat{\rho}]+\sum_{\bm{x}\in \Lambda}\left(\hat{L}_{\bm{x}} \hat{\rho} \hat{L}_{\bm{x}}^{\dagger}-\frac{1}{2}\left\{\hat{L}_{\bm{x}}^{\dagger} \hat{L}_{\bm{x}}, \hat{\rho}\right\}\right).
\label{eq:gksl_eq}
\end{equation}
Here, $\mathcal{L}$ is the Liouvillian superoperator acting on the density matrix $\hat{\rho}$, $\hat{H}$ is the Hamiltonian, and $\hat{L}_{\bm{x}}$ are the Lindblad operators.
The Hamiltonian is given by the SU($N$) Fermi-Hubbard model:
\begin{align}
    \hat{H} &=\hat{H}_{\mathrm{hop}}+\hat{H}_{\mathrm{int}}, \\
    \hat{H}_{\mathrm{hop}} &=-J\sum_{\bm{x} \in \Lambda}\sum_{ { {\mu=1}}}^{ { {d}}} \sum_{ { {\sigma=1}}}^{ { {N}}} 
    \left(\hat{c}_{\bm{x},  {\sigma}}^{\dagger} \hat{c}_{\bm{x} +\bm{e}_\mu,  {\sigma}}+\text{H.c.}\right),
    \label{eq:hamhop}\\
    \hat{H}_{\mathrm{int}} &=U \sum_{\bm{x} \in \Lambda} \sum_{  {1\leq \sigma<\tau\leq N}} \hat{n}_{\bm{x},  {\sigma}}\hat{n}_{\bm{x},  {\tau}},
    \label{eq:hamint}
\end{align}
where $\bm{e}_\mu$ is the unit vector in the direction $\mu$ $(=1,\ldots, d)$, and $\hat{H}_{\mathrm{hop}}$ represents the nearest neighbor hopping term. Here, $J\in \mathbb{R}$ is the hopping amplitude and $U\in \mathbb{R}$ is the strength of interaction.
The Lindblad operator 
\begin{equation}
    \hat{L}_{\bm{x}}=\sqrt{2\gamma}\sum_{1\leq\sigma<\tau\leq N}\hat{c}_{\bm{x},\sigma}\hat{c}_{\bm{x},\tau}
\end{equation}
describes a two-body loss at site $\bm{x}$ with rate $\gamma>0$.

\medskip

\subsection{Diagonalization of the Liouvillian}
Generally speaking, it is more difficult to diagonalize a Liouvillian than a Hamiltonian of a closed system. However, in our case, the problem simplifies to the diagonalization of an effective non-Hermitian Hamiltonian~\cite{torres_closed-form_2014,yoshida_fate_2020,nakagawa_exact_2021}.
We first decompose the Liouvillian as $\mathcal{L}=\mathcal{K}+\mathcal{J}$ with $\mathcal{K} \hat{\rho}:=-i\left(\hat{H}_{\mathrm{eff}} \hat{\rho}-\hat{\rho} \hat{H}_{\mathrm{eff}}^{\dagger}\right)$, $\mathcal{J} \hat{\rho}:=\sum_{\bm{x} \in \Lambda} \hat{L}_{\bm{x}} \hat{\rho} \hat{L}_{\bm{x}}^{\dagger}$, where $\hat{H}_{\mathrm{eff}}:=\hat{H}-\frac{i}{2} \sum_{\bm{x} \in \Lambda} \hat{L}_{\bm{x}}^{\dagger} \hat{L}_{\bm{x}}$ is the effective non-Hermitian Hamiltonian. 
The explicit form of $\hat{H}_{\mathrm{eff}}$ is given by 
\begin{align}
    \hat{H} _{\mathrm{eff}}&=\hat{H}_{\mathrm{hop}}+\hat{H}_{\mathrm{int}}^\prime, \\
    \hat{H}_{\mathrm{hop}} &=-J\sum_{\bm{x} \in \Lambda}\sum_{ { {\mu=1}}}^{ { {d}}} \sum_{ { {\sigma=1}}}^{ { {N}}} 
    \left(\hat{c}_{\bm{x},  {\sigma}}^{\dagger} \hat{c}_{\bm{x} +\bm{e}_\mu,  {\sigma}}+\text{H.c.}\right),
        \label{eq:hamhopeff}\\
    \hat{H}_{\mathrm{int}}^\prime &=(U-i\gamma) \sum_{\bm{x} \in \Lambda} \sum_{  {1\leq \sigma<\tau\leq N}} \hat{n}_{\bm{x},  {\sigma}}\hat{n}_{\bm{x},  {\tau}}.
    \label{eq:haminteff}
\end{align}

Since $\hat{H}_{\mathrm{eff}}$ is non-Hermitian, 
left and right eigenvectors
are no longer related by complex conjugation. 
Throughout this paper, we use the right eigenvectors unless otherwise stated. Since $\hat{H}_{\mathrm{eff}}$ conserves the number of particles, we can choose simultaneous eigenstates of 
$\hat{H}_{\mathrm{eff}}$ and 
$\hat{N}_f$.
Let $\ket{n,a}$ be such that 
$\hat{H}_{\mathrm{eff}}\ket{n,a}=E_{n,a}\ket{n,a}$
and $\hat{N}_f \ket{n,a}=n\ket{n,a}$. 
Here, $a$ labels different states with fixed particle number $n$.
Then one can diagonalize $\mathcal{K}$ as $\mathcal{K}\hat{\rho}_{a,b}^{(n,m)}=\lambda_{a,b}^{(n,m)}\hat{\rho}_{a,b}^{(n,m)}$, where 
$\hat{\rho}_{a,b}^{(n,m)}=\ket{n,a}\bra{n+m,b}$, $\lambda_{a,b}^{(n,m)}=-i(E_{n,a}-E^*_{n+m,b})$. Since the superoperator $\mathcal{J}$ always decreases the number of particles, the matrix representation of $\mathcal{J}$ in the basis $\{\hat{\rho}_{a,b}^{(n,m)}\}_{n,m,a,b}$ is block-triangular when the rows and columns are properly ordered.
Thus, the eigenvalues of $\mathcal{L}$ are given by $\lambda_{a,b}^{(n,m)}$.

\medskip

\subsection{Steady states}
A steady state of the GKSL master equation is defined as a state $\hat{\rho}$ such that $\mathcal{L}\hat{\rho}=0$. 
To find a steady state, it is convenient to use $\hat{H}_{\mathrm{eff}}$, because when $\ket{\psi}$ is an eigenstate of $\hat{H}_{\mathrm{eff}}$ with real eigenvalue, $\ket{\psi}\bra{\psi}$ is a steady state~\cite{nakagawa_exact_2021}. 

To construct $\ket{\psi}$, we write by $\{v_j({\bm{x}})\}_{\bm{x}\in\Lambda}$ the $j$th normalized eigenvector of $\hat{H}_{\mathrm{hop}}$ in the one-particle Hilbert space. Now we define a new set of operators,
\begin{equation}
 \hat{a}^\dagger_{j,\sigma} = \sum_{\bm{x}\in\Lambda} v_j ({\bm{x}}) \hat{c}^\dagger_{\bm{x},\sigma} \quad (j=1,\ldots,L^d),
\end{equation}
and we denote by $\ket{0}$ the normalized vacuum state annihilated by all $\hat{c}_{\bm{x},\sigma}$. Then, a Slater determinant $\hat{a}^\dagger_{j_1,1} \ldots \hat{a}^\dagger_{j_n,1}|{0}\rangle$ is an eigenstate of $\hat{H}_{\mathrm{hop}}$. Since $\hat{H}_{\mathrm{hop}}$ is Hermitian, the eigenvalue is real. Furthermore, since this state has no double occupancy due to the Pauli exclusion principle, it is also an eigenstate of $\hat{H}_{\mathrm{int}}^\prime$ with eigenvalue zero. Therefore, it is an eigenstate of $\hat{H}_{\mathrm{eff}}$ with real eigenvalue. 

Next, we define spin raising and lowering operators as
\begin{equation}
\hat{F}^{\sigma,\tau}=\sum_{\bm{x}\in\Lambda}\hat{c}_{\bm{x},\sigma}^\dagger \hat{c}_{\bm{x},\tau} \quad (\sigma\neq \tau).
\end{equation}
Then, $\hat{F}^{\sigma,\tau}$ commute with $\hat{H}_{\mathrm{eff}}$. Therefore, 
\begin{gather}
    \ket{\psi}=\left(\hat{F}^{N,1}\right)^{M_N} \cdots \left(\hat{F}^{2,1}\right)^{M_2} \hat{a}^\dagger_{j_1,1} \ldots \hat{a}^\dagger_{j_n,1}|{0} \rangle,    \label{eq:fully_poralized}\\
    (0\leq M_\sigma,\ M_2+\ldots+M_N\leq n)
    \nonumber
\end{gather}
forms a family of eigenstates of $\hat{H}_{\mathrm{eff}}$ with real eigenvalue. These steady states are 
generalizations of the steady states in the SU(2)~\cite{foss-feig_steady-state_2012,sponselee_dynamics_2018,nakagawa_exact_2021} and SU(3) ~\cite{rosso_eightfold_2022} Hubbard model with two-body loss. 

\medskip
\section{Liouvillian gap}
\subsection{Definition}
The Liouvillian gap $g$ is given by $g=-\Re[\Delta \lambda]$, where $\Delta \lambda$ is a nonzero eigenvalue of $\mathcal{L}$ whose real part is closest to zero. 
When the eigenvalues of the Liouvillian are given by $\lambda_{a,b}^{(n,m)}=-i(E_{n,a}-E^*_{n+m,b})$, the 
$g$ is given by $g=-\Im [\Delta E]$, where $\Delta E$ is a nonzero eigenvalue of $\mathcal{\hat{H}_{\mathrm{eff}}}$ whose imaginary part is closest to zero. 
\medskip

\subsection{Calculation of the Liuouvillian gap}
\begin{figure}[H]
 \centering
  \includegraphics[width=0.875\linewidth]{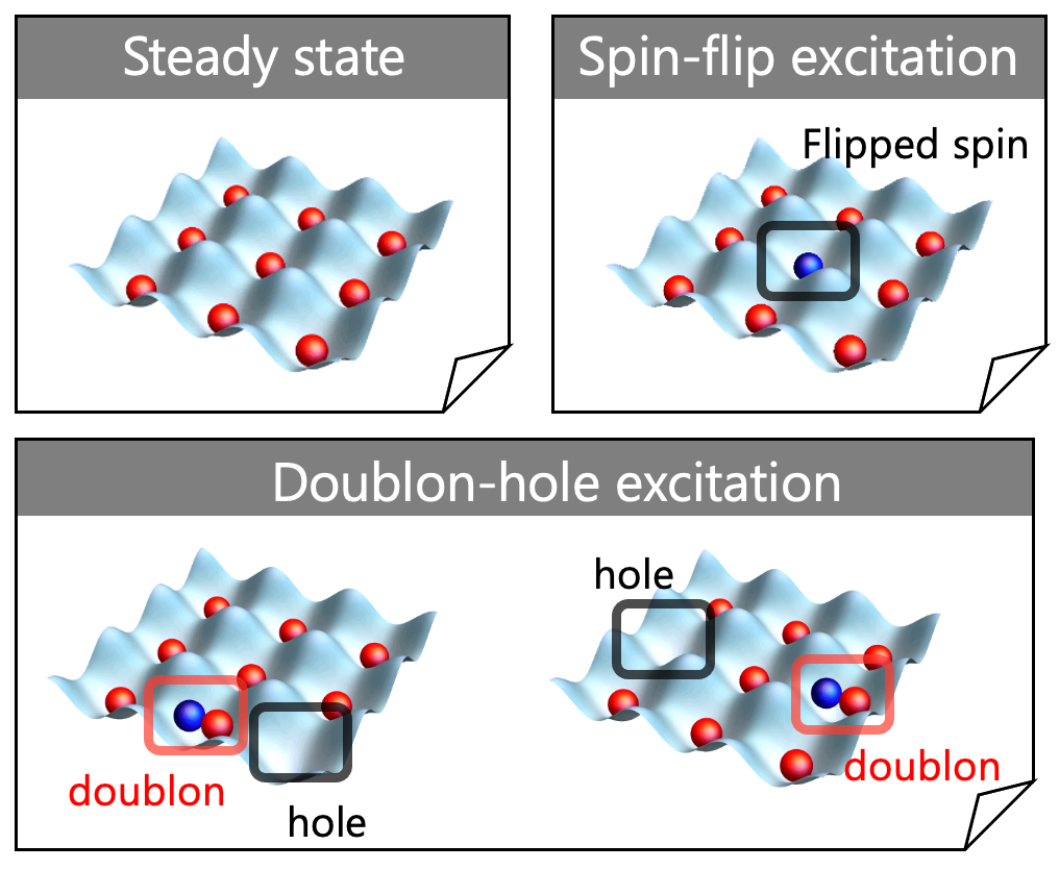}
  \caption{A schematic picture of the steady state \eqref{eq:steady_state} and the spin-flip and doublon-hole excitations from the steady state \eqref{eq:ansatz1}. Blue and red (black and gray) balls represent fermions with spin $\sigma$ and $\tau$, respectively.
  }
  \label{fig:doublon-hole}
\end{figure}

Here we calculate the analytical form of the Liouvillian gap. In the thermodynamic limit, the spectrum of $\mathcal{\hat{H}_{\mathrm{eff}}}$ is gapless along the imaginary axis when $n_f \leq 1$ and gapped when $n_f > 1$~\footnote{This is because when $n_f > 1$, there must be a double occupancy, and thus $-\Im [\Delta E]\geq \gamma$.}. We concentrate on the marginal case $n_f = 1$. In this sector, by applying spin raising and lowering operators, the steady state of the form \eqref{eq:fully_poralized} can be transformed to 
\begin{equation}
    \ket{\mathrm{FM}_{\tau}} = \prod_{j=1}^{L^d}\hat{a}^\dagger_{j,1}|{0} \rangle=
    e^{i\theta}\prod_{\bm{x} \in \Lambda} {\hat c}^\dagger_{\bm{x}, \tau} \ket{0},
    \label{eq:steady_state}
\end{equation}
where $e^{i\theta}$ is a determinant of the unitary matrix $\{v_j ({\bm{x}})\}_{j,\bm{x}}$, and without loss of generality we set $e^{i\theta}=1$ in the following. We have checked numerically for small system sizes that $\Delta E$ is given by the eigenstates of the form 
\begin{equation}
    \ket{\psi}=\sum_{\bm{x}, \bm{y} \in \Lambda}g(\bm{x}, \bm{y}) \hat{c}_{\bm{x}, \sigma}^{\dagger} \hat{c}_{\bm{y}, \tau}\ket{\mathrm{FM}_\tau},
    \label{eq:ansatz1}
\end{equation}
where $g(\bm{x}, \bm{y})$ is a complex-valued function of $\bm{x}$ and $\bm{y}$.
These states are superpositions of the spin-flip and doublon-hole excitations from the steady state as illustrated in Fig.~\ref{fig:doublon-hole}.

Then we solve the eigenequation $\hat{H}_{\mathrm{eff}}\ket{\psi}=E \ket{\psi}$, extending the standard method in analyzing the two-body problem in the Hermitian Fermi-Hubbard model~\cite{song_pseudospin_1991,essler2005}. By a direct calculation, one obtains the following commutation relation:
\begin{align}
\left[\hat{H}_{\mathrm{eff}},\hat{c}_{\bm{x}, \sigma}^{\dagger} \hat{c}_{\bm{y}, \tau}\right] \!=\!
    &-J\sum_{\mu=1}^d\sum_{s=\pm}\left[\hat{c}_{\bm{x}
    +s\bm{e}_\mu, \sigma}^{\dagger} \hat{c}_{\bm{y}, \tau} \!-\! \hat{c}_{\bm{x}, \sigma}^{\dagger} \hat{c}_{\bm{y}
    +s\bm{e}_{\mu},\tau}\right] \nonumber\\
    &+
    (U-i\gamma)(1-\delta_{\bm{x},\bm{y}})\hat{c}_{\bm{x}, \sigma}^{\dagger} \hat{c}_{\bm{y}, \tau}.
\end{align}
By substituting this and $\hat{H}_{\mathrm{eff}} \ket{\mathrm{FM}_\tau}=0$ into $\hat{H}_{\mathrm{eff}}\ket{\psi}=E \ket{\psi}$ and comparing the coefficients, one obtains 
\begin{align}
    -J\sum_{\mu=1}^d \sum_{s=\pm}&\left[g(\bm{x}+s \bm{e}_{\mu}, \bm{y})-g(\bm{x}, \bm{y}+s \bm{e}_{\mu})\right]\nonumber\\
    &= [E-(U-i\gamma)(1-\delta_{\bm{x},\bm{y}})]g(\bm{x}, \bm{y}).
    \label{eq:two_body}
\end{align}
with periodic boundary conditions
${g(\bm{x}+L\bm{e}_{\mu},\bm{y})}={g(\bm{x},\bm{y}+L\bm{e}_{\mu})}={g(\bm{x},\bm{y})}$. This equation can be separated into those of center-of-mass and relative coordinates $\bm{m}=(\bm{x}+\bm{y})/2,$ $\bm{n}=\bm{x}-\bm{y}$. Due to the translational symmetry, the center-of-mass momentum $\bm{k}$ is conserved. Thus, we write 
\begin{equation}
   g(\bm{x}, \bm{y})=\sum_{\bm{k}} e^{i \bm{k} \cdot \bm{m}} f_{\bm{k}}(\bm{n}),
   \label{eq:ansatz2}
\end{equation}
where $\bm{k}=\left(k^{1}, \ldots, k^{d}\right)$ and each $k^{\mu}$ is quantized as $k^{\mu}=\frac{2 \pi l_{\mu}}{L}\left(l_{\mu}=1, \ldots, L\right)$ due to the periodic boundary conditions.
Note that the boundary conditions for $f_{\bm{k}}(\bm{n})$ are $f_{\bm{k}}(\bm{n}+L\bm{e}_\mu)=(-1)^{l_\mu} f_{\bm{k}}(\bm{n})$. By substituting Eq.~\eqref{eq:ansatz2} into Eq.~\eqref{eq:two_body} and writing $E$ as $E_{\bm{k}}$, we have
\begin{align}
\sum_{\mu=1}^d  &[-i r_{\bm{k}}^\mu f_{\bm{k}}(\bm{n}+\bm{e}_\mu)+i r_{\bm{k}}^\mu f_{\bm{k}}(\bm{n}-\bm{e}_\mu)]\nonumber \\
&=\left[E_{\bm{k}}-u\left(1-\delta_{\bm{n},\bm{0}}\right)\right]f_{\bm{k}}(\bm{n}),
\label{eq:tb_imp_pre}
\end{align}
where $r_{\bm{k}}^\mu=2J\sin (k^\mu/2)$ and $u=U-i\gamma$. Finally, by setting $f_{\bm{k}}(\bm{n})\mapsto i^{\left(\sum_{\mu=1}^d n_\mu\right)}f_{\bm{k}}(\bm{n})$, it can be rewritten as
\begin{equation}
    \sum_{\mu=1}^d r_{\bm{k}}^\mu [f_{\bm{k}}(\bm{n}+\bm{e}_\mu)+f_{\bm{k}}(\bm{n}-\bm{e}_\mu)]
=\left[E_{\bm{k}}-u\left(1-\delta_{\bm{n},\bm{0}}\right)\right]f_{\bm{k}}(\bm{n}).
\label{eq:tb_imp}
\end{equation}
The boundary conditions for $f_{\bm{k}} (\bm{n})$ read $f_{\bm{k}}(\bm{n}+L\bm{e}_\mu)=(-i)^{L+2l_\mu} f_{\bm{k}}(\bm{n})$. Equation~\eqref{eq:tb_imp} can be regarded as a Schr\"odinger equation for a tight-binding model with a complex-valued impurity at the origin $\bm{0}$.

Next, we solve the eigenequation \eqref{eq:tb_imp}. For simplicity, we assume the periodic boundary conditions, i.e., we consider the case where $L+2l_\mu$ is a multiple of four, but the other cases can be treated similarly.
We first expand $f_{\bm{k}}(\bm{n})$ as $f_{\bm{k}}(\bm{n})=\frac{1}{L^d}\sum_{\bm{\waveq}}e^{i \bm{\waveq} \cdot \bm{n}} F_{\bm{k}}(\bm{\waveq})$, where $F_{\bm{k}}(\bm{\waveq})$ is the Fourier transform of $f_{\bm{k}}(\bm{n})$. Substituting this into
Eq.~\eqref{eq:tb_imp}, using $\delta_{\bm{n},\bm{0}}=\frac{1}{L^d}\sum_{\bm{\waveq}}e^{i \bm{\waveq}\cdot\bm{n}}$, and comparing the coefficients of $e^{i \bm{\waveq}\cdot\bm{n}}$ on both sides, one obtains
\begin{equation}
    \left[ E_{\bm{k}}-u -\varepsilon_{\bm{k}}(\bm{\waveq})\right]F_{\bm{k}}(\bm{\waveq})=-\frac{u}{N}\sum_{\bm{\waveq}^\prime}F_{\bm{k}}(\bm{\waveq}^\prime),
    \label{eq:sum_equation}
\end{equation}
where $\varepsilon_{\bm{k}}(\bm{\waveq})=\sum_{\mu}2r_\mu \cos \waveq_\mu$. 
By writing $S_{\bm{k}}=\sum_{\bm{\waveq}}F_{\bm{k}}(\bm{\waveq})$,
this equation can be rewritten as 
\begin{equation}
    S_{\bm{k}}=-\frac{1}{L^d}\sum_{\bm{\waveq}}\frac{u}{E_{\bm{k}}-u -\varepsilon_{\bm{k}}(\bm{\waveq})} S_{\bm{k}},
\end{equation}
which has a nontrivial solution only if
\begin{equation}
   \frac{1}{L^d}\sum_{\bm{\waveq}}\frac{u}{E_{\bm{k}}-u -\varepsilon_{\bm{k}}(\bm{\waveq})} =-1.
      \label{eq:self_consistent_sum}
\end{equation}
In the limit $L\to\infty$, Eq.~\eqref{eq:self_consistent_sum} can be rewritten as 
\begin{equation}
   \frac{1}{\pi^d}\int_{0}^{\pi} d^d \waveq\frac{u}{E_{\bm{k}}-u -\sum_{\mu=1}^d 2r_{\bm{k}}^\mu \cos \waveq_\mu }
    =-1.
   \label{eq:self_consistent_int}
\end{equation}
Since we are interested in the Liouvillian gap, we concentrate on the modes with small $-\Im[E_{\bm{k}}]$, which is realized by small $\bm{k}$. In this case, $r_{\bm{k}}^\mu=2J\sin (k^\mu/2)$ implies that the condition $|u|\gg \sum_{\mu=1}^d r_{\bm{k}}^\mu$ is satisfied when $L$ is large enough. Thus we assume that $|u|\gg \sum_{\mu=1}^d r_{\bm{k}}^\mu$. Assuming further that $|E_{\bm{k}}|\ll|u|$,
we have
\begin{equation}
    E_{\bm{k}}=
    -\frac{8J^2\sum_{\mu=1}^d\sin^2(k^\mu/2)}{u}.
    \label{eq:pre_liouvillian_gap}
\end{equation}
This solution actually satisfies $|E_{\bm{k}}|\ll|u|$. The imaginary part of $E_{\bm{k}}$ is closest to zero when $k^{1}=2\pi/L$ and $k^{2}=\ldots=k^{d}=0$. Thus, we arrive at the explicit expression for the Liouvillian gap: 
\begin{equation}
g
=\frac{8J^2\gamma \sin^2(\pi/L)}{U^2+\gamma^2}
=\frac{8\pi^2J^2\gamma }{(U^2+\gamma^2)L^2}+O(L^{-3}).
\label{eq:liouvillian_gap}
\end{equation}
This is the central result of this paper. Clearly, the Liouvillian gap is proportional to $1/L^2$ for large $L$, but does not depend on $N$ or $d$. 
This result is consistent with the result 
for the one-dimensional model obtained by the Bethe ansatz (See Eq.~(5) in Ref.~\cite{nakagawa_exact_2021}). 
We also discuss in Appendix \ref{sec:one-dimension} that in one dimension, the energy of the bound state can be calculated without assuming that $|u|\gg \sum_{\mu=1}^d r_{\bm{k}}^\mu~$.

\medskip

\section{Dynamics of a single spin flip}
The analysis of the spin-flip and doublon-hole excitation from the steady state also applies to the dynamics of a single spin flip. We consider a pure state with a single spin flip
\begin{equation}
    \ket{\psi_0}=\hat{c}_{\bm{0}, \sigma}^{\dagger} \hat{c}_{\bm{0}, \tau}\ket{\mathrm{FM}_\tau}\quad (\sigma\neq \tau).
    \label{eq:init_condition}
\end{equation}
In this section, we calculate the dynamics of the survival probability of a spin flip at time $t$:  $\mathcal{N}(t):=\Tr\left[\hat{N}_\sigma e^{\mathcal{L}t}\hat{\rho}_0\right]=\Tr\left[\hat{N}_\sigma e^{\mathcal{L}t}\left(\ket{\psi_0}\bra{\psi_0}\right)\right]$ using the Lindblad equation \eqref{eq:gksl_eq}. In our setting, $\mathcal{N}(t)$ 
reads (see Appendix \ref{sec:num_spin_flip})
\begin{equation}
   \mathcal{N}(t)=\braket{\psi(t)}
   \label{eq:num_spin_flip}
\end{equation}
where $\ket{\psi(t)}=e^{-i\hat{H}_\mathrm{eff}t}\ket{\psi_0}$ can be calculated using the method described earlier. To proceed, we first note that the initial condition \eqref{eq:init_condition} corresponds to Eq.~\eqref{eq:ansatz1} with $g(\bm{x},\bm{y})=\delta_{\bm{x},\bm{0}}\delta_{\bm{y},\bm{0}}$, which can be expanded as Eq.~\eqref{eq:ansatz2} with $f_{\bm{k}}(\bm{n})=\frac{1}{L^d} \delta_{\bm{n},\bm{0}}$. For each center-of-mass momentum $\bm{k}$, we denote by $E_{\bm{k},j}$ and $f_{\bm{k},j}(\bm{n})$ the $j$th eigenvalue and the corresponding eigenvector of Eq.~\eqref{eq:tb_imp}. In the following, we assume that $f_{\bm{k},j}(\bm{n})$ $(j=1,\ldots, L^d)$ form a complete basis~\footnote{Note that this is not always the case, since the Hamiltonian is non-Hermitian. For example, in a one-dimensional lattice with $L=2$ and $k=\pi$, equation \eqref{eq:tb_imp} reads 
\begin{equation*}
\begin{pmatrix}
0 & 4J \\
4J & u \\
\end{pmatrix}
\begin{pmatrix}
f_\pi(0) \\
f_\pi(1) \\
\end{pmatrix}
=
E_\pi
\begin{pmatrix}
f_\pi(0) \\
f_\pi(1) \\
\end{pmatrix}
\end{equation*}
and the matrix in this equation is not diagonalizable when $u=\pm8iJ$ and $J\neq0$.}. When $\delta_{\bm{n},\bm{0}}$ is expanded as $\delta_{\bm{n},\bm{0}}=\sum_{j}d_{\bm{k},j} f_{\bm{k},j} (\bm{n})$ with coefficients $d_{\bm{k},j}$, 
$\ket{\psi(t)}$ and $\mathcal{N}(t)$ read
\begin{gather}
    \ket{\psi(t)}=
    \sum_{\bm{x}, \bm{y} \in \Lambda}g(\bm{x}, \bm{y}, t) \hat{c}_{\bm{x}, \sigma}^{\dagger} \hat{c}_{\bm{y}, \tau}\ket{\mathrm{FM}_\tau},
    \label{eq:wave_function}\\
    g(\bm{x}, \bm{y}, t)=\frac{1}{L^d}\sum_{\bm{k},j} d_{\bm{k},j} e^{-i E_{\bm{k},j} t} e^{i \bm{k} \cdot (\bm{x}+\bm{y})/2} f_{\bm{k},j}(\bm{x}-\bm{y}),
    \label{eq:wave_function2}\\
    \mathcal{N}(t)=\frac{1}{L^d}\sum_{\bm{k},\bm{n},j,l}d_{\bm{k},j} d^*_{\bm{k},l} e^{-i(E_{\bm{k},j}-E_{\bm{k},l}^*)t}f_{\bm{k},j}(\bm{n})f^*_{\bm{k},l}(\bm{n}).
    \label{eq:particle_number}
\end{gather}
Note that $\sum_{\bm{n}}f_{\bm{k},j}(\bm{n})f^*_{\bm{k},l}(\bm{n})\neq0$ for $j\neq l$ in general because the Hamiltonian is non-Hermitian. We prove in Appendix \ref{sec:proof_ss} that
\begin{gather}
    \lim_{t\to\infty}\ket{\psi(t)}=\frac{1}{L^d}\sum_{\bm{x}\in \Lambda} \hat{c}_{\bm{x}, \sigma}^{\dagger} \hat{c}_{\bm{x},\tau}\ket{\mathrm{FM}_\tau},
    \label{eq:psi_ss}\\
    \lim_{t\to\infty} \mathcal{N}(t)=\frac{1}{L^d},
    \label{eq:n_ss} 
\end{gather}
and therefore $\lim_{t\to\infty} \mathcal{N}(t)=0$ in the thermodynamic limit. In the following, we calculate $\mathcal{N}(t)$ analytically in two limits: strongly and weakly interacting and dissipative limit. 
We also present numerical results of $\mathcal{N}(t)$ for various values of the parameters.
\medskip

\subsection{Strongly interacting and dissipative limit} 
In this limit, Eq.~\eqref{eq:sum_equation} can be approximated as $F_{\bm{k}}(\bm{\waveq})\simeq\frac{1}{N}\sum_{\bm{\waveq}^\prime}F_{\bm{k}}(\bm{\waveq}^\prime)$, which means that $F_{\bm{k}}(\bm{\waveq})$ is almost constant 
as a function of $\bm{\waveq}$. Thus, 
$f_{\bm{k}}(\bm{n})=\frac{1}{L^d}\sum_{\bm{\waveq}}e^{i \bm{\waveq} \cdot \bm{n} }F_{\bm{k}}(\bm{\waveq})$ can be approximated as $f_{\bm{k}}(\bm{n})=\frac{1}{L^d} \delta_{\bm{n},\bm{0}}$, and therefore
\begin{align}
    \mathcal{N}(t)&
    =\frac{1}{L^d}\sum_{\bm{k}}e^{-2\Gamma t\sum_{\mu=1}^d \sin^2 (k^\mu/2)} \\
    &\to \left(\int_{0}^1 e^{-2\Gamma t \sin^2(\pi x)}dx\right)^d = \left[e^{-\Gamma t} I_0(\Gamma t)\right]^d,
    \label{eq:spin_impurity_strong_bessel}
\end{align}
where $\Gamma=\frac{8J^2\gamma}{U^2+\gamma^2}$, $I_0$ is the modified Bessel function of the first kind, and we took the limit $L\to\infty$ in the second 
line. 
Since the asymptotic form of $I_0(x)$ in the $x\to \infty$ limit is $I_0(x)\simeq e^x/\sqrt{2\pi x}$, 
$\mathcal{N}(t)$ approaches $\mathcal{N}(t)\simeq (2\pi \Gamma t)^{-d/2}$ 
when $t$ is sufficiently larger than $\Gamma^{-1}$.
The new time scale $\Gamma^{-1}$ is related to the continuous quantum Zeno effect~\cite{syassen_strong_2008,zhu_suppressing_2014,nakagawa_dynamical_2020, rosso_dynamical_2023}.
When $0\leq\gamma/U\leq1$, $\Gamma$ increases monotonically with $\gamma$, but when $\gamma/U\geq1$, $\Gamma$ decreases monotonically with $\gamma$. This is due to the continuous quantum Zeno effect, by which the double occupancy necessary for particle loss is prohibited. It is also interesting to compare our results with previous studies. 
In Ref.~\cite{rosso_dynamical_2023}, the authors numerically study the dynamics of the one-dimensional Fermi-Hubbard model with strong interaction or strong dissipation for N\'{e}el and maximally mixed initial states, and found the power-law scaling of the particle number: $t^{-1/3}$ ($t^{-1/4}$) for the N\'{e}el (maximally-mixed) initial conditions. 

\medskip

\subsection{Weakly interacting and dissipative limit} 
In this limit, $\mathcal{N}(t)$ exhibits an exponential decay with time, i.e., $\mathcal{N}(t)\simeq e^{-2\gamma t}$. The reason is as follows. First, when $\gamma$ is small, $f_{\bm{k},j}$ can be taken to satisfy
$\sum_{\bm{n}}f_{\bm{k},j}(\bm{n})f^*_{\bm{k},l}(\bm{n})\simeq \delta_{j,l}$ for $j\neq l$. Thus, Eq. \eqref{eq:particle_number} reduces to
\begin{equation}
    \mathcal{N}(t)=\frac{1}{L^d}\sum_{\bm{k},j}|d_{\bm{k},j}|^2 e^{2\Im[E_{\bm{k},j}]t}.
\end{equation}
It is easy to see that $\Im[E_{\bm{k},j}]\simeq -\gamma$ for scattering states, because $\left|f_{\bm{k},j}(\bm{0})\right|=O(1/L)$ for these states. Next, we show this for bound states. In the weakly interacting and dissipative limit, $|u|/\sum_{\mu=1}^d r_{\bm{k}}^\mu \ll 1$ for almost all $\bm{k}$. In one dimension
, the energy of the bound state can be approximated as $E_{k}=u\pm2r_k$, and therefore $\Im[E_{\bm{k}}]\simeq -\gamma$. In higher dimensions, we do not have the explicit form of $E_{\bm{k}}$. But one finds $\bm{q}$ such that $|E_{\bm{k}}-u -\sum_{\mu=1}^d 2r_{\bm{k}}^\mu \cos \waveq_\mu|\simeq 0$ mainly contribute to the integral on the left-hand side of Eq.~\eqref{eq:self_consistent_int}. Therefore, $\Im[E_{\bm{k}}]$ must satisfy $\Im[E_{\bm{k}}]\simeq\Im [u]=-\gamma$. 
Thus, we have $\Im[E_{\bm{k},j}]\simeq -\gamma$ for almost all $\bm{k}$ and $j$, which yields the desired result. 

\begin{figure}[H]
 \centering
  \includegraphics[width=\linewidth]{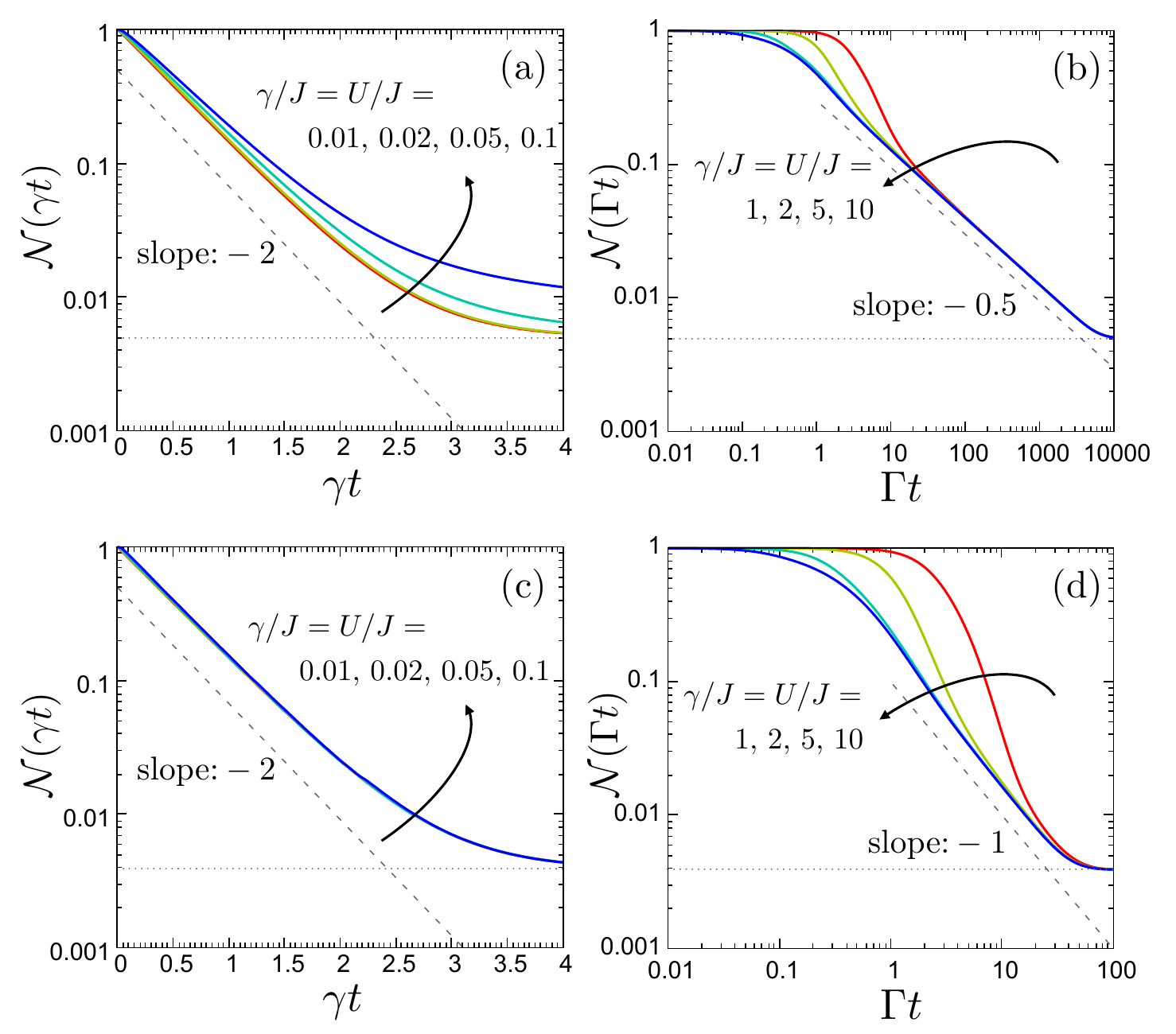}
  \caption{Dynamics of the survival probability of a spin flip in one dimension with $L=200$ (a and b) and in two dimensions with $L=16$ (c and d). In (c), the curves lie almost on top of each other. The dotted lines are guides to the eye. The horizontal dotted line in each panel indicates the survival probability of the spin flip in the steady state~($1/L^d$).}
  \label{fig:flipped_spin}
\end{figure}

\subsection{Intermediate regions} 
Here we 
calculate the dynamics of a single flipped spin by numerical exact diagonalization of Eq.~\eqref{eq:particle_number},
and the result is shown in Fig.~\ref{fig:flipped_spin}. In Figs.~\ref{fig:flipped_spin}(a) and \ref{fig:flipped_spin}(c), the survival probability of the spin flip as a function of the rescaled time
$\gamma t$ is plotted on a logarithmic scale when $U$ and $\gamma$ are smaller than $J$. From these plots, one finds that the dynamics of the particle number approaches an exponential decay as $U/J$ and $\gamma/J$ get smaller. Although the slope is predicted to be $-2$ in the limit $L\to\infty$, 
the numerical results are larger than this value because of the finite-size effect. In Figs.~\ref{fig:flipped_spin}(b) and \ref{fig:flipped_spin}(d), the survival probability of the spin flip as a function of the rescaled time $\Gamma t$ is plotted on a log-log scale when $U$ and $\gamma$ are larger than $J$. Now we see that the curve approaches Eq.~\eqref{eq:spin_impurity_strong_bessel} as $U/J$ and $\gamma/J$ get larger, and thus asymptotically approaches a power law for large $\Gamma t$.

\subsection{Effect of the boundary conditions}
\label{sec:obc}
\begin{figure}[H]
 \centering
  \includegraphics[width=0.85\linewidth]{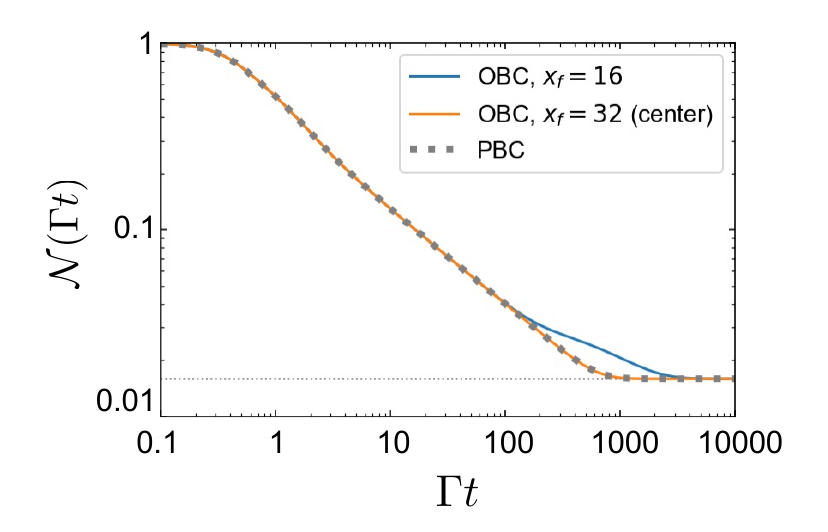}
  \caption{Dynamics of the survival probability of a spin flip with open boundary conditions in one dimension $(L=63)$. The position of the initial spin flip is denoted by $x_f$. The horizontal dotted line indicates the survival probability of the spin flip in the steady state.}
  \label{fig:open_boundary}
\end{figure}

Here we consider the effect of the edges. For simplicity, we consider open boundary conditions and ignore the trapping potentials. This means a sharp edge, which was realized in a recent experiment using a digital micromirror device~\cite{sompet_realizing_2022}.
We solve Eq.~\eqref{eq:two_body} with open boundary conditions (OBC) in one dimension, and the result is shown in Fig.~\ref{fig:open_boundary}. Here we consider the strong interaction and dissipation case $(U/t=\gamma/t=4)$. If the position of the spin flip $x_f$ is located at the center of the chain ($x_f=32$ in Fig.~\ref{fig:open_boundary}), the dynamics is identical to the case of the periodic boundary conditions (PBC). When $x_f$ is not at the center of the chain ($x_f=16$ in Fig.~\ref{fig:open_boundary}), even though the late-time dynamics differs from the case of the PBC, the early- and intermediate-time dynamics agree well with those under PBC.

\medskip

\section{Conclusion and outlook}
We have analyzed spin-flip and doublon-hole excitations from a ferromagnetic steady state 
of the SU($N$) Fermi-Hubbard model with a two-body loss on a $d$-dimensional hypercubic lattice and obtained 
(i) the analytical form of the Liouvillian gap, and (ii) dynamics of a single-spin flip in a ferromagnetic environment. As a result, at $1/N$ filling, we found that the Liouvillian gap of the system with linear size $L$ is proportional to $1/L^2$ but does not depend on $d$ or $N$. We also showed that by decreasing the interaction and dissipation, a crossover from the power-law to the exponential decay occurs.

In this paper, we focused on a hypercubic lattice, but our methods can be applied to more general lattices. 
We also note that the dynamics of a single spin flip 
can be calculated with a low computational cost 
even in the absence of translational symmetry as shown in Sec. \ref{sec:obc}, thus our method can be applied to the case with confinement potentials.

Moreover, when $N=2$ and the lattice is bipartite, the model discussed in this paper can be mapped to a dissipative Hubbard model subject to spontaneous emission~\cite{nakagawa_eta_2021} by the Shiba transformation~\cite{shiba_thermodynamic_1972}. Thus, our approach is applicable to the model with appropriate identification of the filling and the interaction parameter $U$. 
Our approach is also valid for a tight binding model on a bipartite lattice with dephasing noise~\cite{medvedyeva_exact_2016}, because the model can be mapped to a Hubbard model with imaginary-interaction strength.

We expect that our findings can be tested experimentally with a site- and spin-resolved quantum gas microscope~\cite{bakr_quantum_2009,cheuk_quantum-gas_2015,haller_single-atom_2015,yamamoto_ytterbium_2016,gross_quantum_2017} of ultracold alkaline-earth-like atoms in optical lattices.

\medskip
\begin{acknowledgments}
We thank Masaya Nakagawa, Kazuki Yamamoto, Yoshiro Takahashi, Yosuke Takasu, and Kantaro Honda for valuable discussions. H.K. was supported by MEXT KAKENHI Grant-in-Aid for Transformative Research Areas A ``Extreme Universe" No. JP21H05191, JSPS KAKENHI Grant No. JP18K03445, and the Inamori Foundation. 
H.Y. was supported by JSPS KAKENHI Grant-in-Aid for JSPS fellows Grant No. JP22J20888 and JSR Fellowship, the University of Tokyo.

\end{acknowledgments}

\appendix

\section{Bound state in one dimension}
\label{sec:one-dimension}
In the main text, we analyzed Eq.~\eqref{eq:self_consistent_int} in the limit $|u|\gg \sum_{\mu=1}^d r_{\bm{k}}^\mu$. Here we solve the equation analytically in one dimension. In one dimension, we write $\bm{k}$, $\bm{\waveq}$ and $r^{\mu=1}$ as $k$, $\waveq$ and $r$. Then, the following integral with a complex parameter $z$
\begin{equation}
   D(z)= \frac{1}{\pi}\int_0^\pi d\waveq \frac{1}{z- \cos\waveq}
   \label{eq:int_n0}
\end{equation}
converges to 
\begin{equation}
   D(z)= \frac{i (-1)^{\lfloor\frac{\arg [(1-z)/(1+z)]}{2\pi}\rfloor}}{\sqrt{1-z^2}}
\end{equation}
when $|z|\geq 1$ or $z\in \mathbb{R}$. Here, $\lfloor x \rfloor$ is the floor function and $\sqrt{x}$ is the principal value of the square root of $x$. Substituting it into Eq.~\eqref{eq:self_consistent_int} and taking the square of both sides, we have 
\begin{equation}
   \frac{u^2}{(E_k-u)^2-4r_k^2}=1,
\end{equation}
and thus
\begin{equation}
   E_k=u\pm\sqrt{u^2+4r_k^2}.
  \label{eq:energy}
\end{equation}
We again analyze the limit $|u|\gg r_k$, and check that Eq.~\eqref{eq:energy} reproduces the results in the main text. In this limit, by choosing $E_k=u+\sqrt{u^2+4r_k^2}$ if $U>0$ and $E_k=u-\sqrt{u^2+4r_k^2}$ if $U<0$, we obtain 
\begin{equation}
   E_k=-\frac{2r_k^2}{u}=-\frac{8J^2\sin^2(k/2)}{u},
\end{equation}
which is consistent with Eq.~(\ref{eq:pre_liouvillian_gap}). Next, we calculate the wave functions of the bound states. In the following, we assume that $U>0$ and $E_k=u+\sqrt{u^2+4r_k^2}$. By substituting Eq.~\eqref{eq:energy} into Eq.~\eqref{eq:sum_equation}, one has
\begin{equation}
    F_k(\waveq)\propto \frac{u}{\sqrt{u^2+4r_k^2}-2r_k \cos \waveq},
    \label{eq:wave_k}
\end{equation}
and thus
\begin{align}
    f_k(n)
    &\propto\frac{1}{L}\sum_{\waveq}\frac{ue^{i\waveq n}}{\sqrt{u^2+4r_k^2}-2r_k \cos \waveq} \\
    &\to \frac{1}{2\pi}\int_0^{2\pi}\frac{ue^{i\waveq n}}{\sqrt{u^2+4r_k^2}-2r_k \cos \waveq}d\waveq.
    \label{eq:wave_n}
\end{align}
We first see the qualitative behavior of $f_k(n)$ in the limits $|u|\gg r_k$ and $|u|\ll r_k$. When $|u|\gg r_k$, \eqref{eq:wave_k} is almost independent of $\waveq$, and hence $f_k(n)$ is localized around $n=0$. On the other hand, when $|u|\ll r_k$, since
\begin{equation}
    F_k(\waveq)\propto \frac{u}{\sqrt{u^2+4r_k^2}-2r_k \cos \waveq}\simeq \frac{u}{2r_k (1-\cos \waveq)+\frac{u^2}{2r_k}},
\end{equation}
$F_k(\waveq)$ has a sharp peak at $\waveq=0$, and thus $\left|f_k(n)\right|$ is almost constant. These behaviors are qualitatively the same for higher-dimensional cases, as discussed in the main text. In one dimension, one can further obtain the analytical form of $f_k(n)$. To this end, we evaluate the integral
\begin{equation}
    D(n,z)=\frac{1}{2\pi}\int_{0}^{2\pi}\frac{e^{i\waveq n}}{z-\cos \waveq}d\waveq\quad (n\in\mathbb{Z}).
\end{equation}
Since this integral is the same as Eq.~\eqref{eq:int_n0} for $n=0$ and satisfies $D(n,z)=D(-n,z)$, we focus on the case $n\geq 1$. By replacing $e^{i\waveq}$ with $\xi$, one has
\begin{equation}
    D(n,z)=\frac{1}{2\pi i} \oint_C \frac{\xi^n}{\xi^2 -2z \xi +1}d\xi,
\end{equation}
where $C$ is the contour taken counterclockwise 
around the unit circle in the complex plane. The poles of the integrand are at 
the solutions of $\xi^2 -2z \xi +1=0$: $\xi=z\pm \sqrt{z^2-1}$. 
Since the product of two solutions of $\xi^2 -2z \xi +1=0$ is always $1$, both poles lie on $C$ or one pole lies inside the contour $C$ and the other outside.
The former case is true if and only if $z\in \mathbb{R}$ and $|z|\leq 1$.
In the former case, the integral diverges; in the latter case, if $\xi=z+\sqrt{z^2-1}$ is inside $C$,
\begin{equation}
    D(n,z)=\frac{(z+\sqrt{z^2-1})^n}{\sqrt{z^2-1}},
\end{equation}
and if $\xi=z-\sqrt{z^2-1}$ is inside $C$,
\begin{equation}
    D(n,z)=-\frac{(z-\sqrt{z^2-1})^n}{\sqrt{z^2-1}}.
\end{equation}
Let us return to Eq.~\eqref{eq:wave_n}. The function $f_k(n)$ is obtained by setting $z=\frac{\sqrt{u^2+4r_k^2}}{2r_k}$. We first see that the case where both poles lie on $C$ is excluded, because $\gamma$ has to be zero if we assume that $z\in \mathbb{R}$, but this implies $\left|\frac{\sqrt{u^2+4r_k^2}}{2r_k}\right|>1$. When $U>0$, $\xi=\frac{\sqrt{u^2+4r_k^2}-u}{2r_k}$ is inside $C$, and therefore
\begin{equation}
f_k(n)\propto \left(\frac{\sqrt{u^2+4r_k^2}-u}{2r_k}\right)^{|n|}.
\end{equation}

\section{Derivation of Eq.~(\ref{eq:num_spin_flip})}
\label{sec:num_spin_flip}
Here, we derive Eq.~\eqref{eq:num_spin_flip} in the main text.
By decomposing $\mathcal{L}$ as $\mathcal{L}=\mathcal{K}+\mathcal{J}$, $\mathcal{N}(t)$ can be expanded as 
\begin{equation}
\mathcal{N}(t)=\Tr\left[\hat{N}_\sigma e^{\mathcal{L}t}\hat{\rho}_0\right]=
\sum_{n=0}^\infty \frac{t^n}{n!} \Tr\left[\hat{N}_\sigma (\mathcal{K}+\mathcal{J})^n\hat{\rho}_0\right].
\label{eq:expansion}
\end{equation}
Note that $\mathcal{K}$ $(\mathcal{J})$ always preserves (decreases) the particle number. Since $\Tr\left[\hat{N}_\sigma \hat{\rho}_0\right]=\Tr\left[\hat{\rho}_0\right]=1$, we have $\Tr\left[\hat{N}_\sigma \mathcal{K}^n \hat{\rho}_0\right]=\Tr\left[ \mathcal{K}^n \hat{\rho}_0\right]$ for all $n=0,1,\ldots$ and the other terms appearing in Eq.~\eqref{eq:expansion} such as $\Tr\left[\hat{N}_\sigma \mathcal{J}\mathcal{K} \hat{\rho}_0\right]$ 
equal to $0$.
Thus, 
\begin{equation}
   \mathcal{N}(t)
   =\sum_{n=0}^\infty \frac{t^n}{n!} \Tr\left[\mathcal{K}^n\hat{\rho}_0\right]
   =\Tr\left[e^{\mathcal{K}t}\hat{\rho}_0\right]
   =\braket{\psi(t)},
\end{equation}
where $\ket{\psi(t)}=e^{-i\hat{H}_\mathrm{eff}t}\ket{\psi_0}$.

\medskip

\section{Proof of Eqs.~(\ref{eq:psi_ss}) and (\ref{eq:n_ss})}
\label{sec:proof_ss}
Here we prove Eqs.~(\ref{eq:psi_ss}) and (\ref{eq:n_ss}). First, when $\bm{k}=\bm{0}$, $r_{\bm{k}}^\mu=2J\sin (k^\mu/2)=0$ for all $\mu$. Thus, $f_{\bm{k}=\bm{0}}(\bm{n})=\delta_{\bm{n},\bm{0}}$ and $E_{\bm{k}=\bm{0}}=0$ is a solution of Eq.~\eqref{eq:tb_imp}. On the other hand, when $\bm{k}\neq\bm{0}$, $f_{\bm{k}}(\bm{n})=\delta_{\bm{n},\bm{0}}$ is not a solution of Eq.~\eqref{eq:tb_imp}. In this case, we can prove that $\operatorname{Im}E_{\bm{k}}<0$, and therefore the corresponding mode vanishes in the limit $t\to \infty$. In fact, if $f_{\bm{k}}$ and $E_{\bm{k}}$ is a solution of Eq.~\eqref{eq:tb_imp} and $f_{\bm{k}}$ is normalized as $\sum_{\bm{n}}|f_{\bm{k}}(\bm{n})|^2=1$, $E_{\bm{k}}$ can be written as 
\begin{gather}    
E_{\bm{k}}= T+V,
\end{gather}  
where
\begin{gather} 
T=\sum_{\bm{n}}\sum_{\mu=1}^d r_{\bm{k}}^\mu [f_{\bm{k}}(\bm{n}+\bm{e}_\mu)+f_{\bm{k}}(\bm{n}-\bm{e}_\mu)]f^*_{\bm{k}}(\bm{n}),\\
V=u\sum_{\bm{n}\neq \bm{0}}|f_{\bm{k}}(\bm{n})|^2.
\end{gather}
Since $r_{\bm{k}}^\mu$ is real, $T$ is also real. Noting that $u=U-i\gamma$, we see that $\operatorname{Im}E_{\bm{k}}=-\gamma\sum_{\bm{n}\neq \bm{0}}|f_{\bm{k}}(\bm{n})|^2$. Since $f_{\bm{k}}(\bm{n})=\delta_{\bm{n},\bm{0}}$ is not a solution of Eq.~\eqref{eq:tb_imp}, $f_{\bm{k}}(\bm{n})\neq 0$ for some $\bm{n}\neq \bm{0}$, so $\operatorname{Im}E_{\bm{k}}<0$ for any ${\bm k} \ne {\bm 0}$. 
Therefore, only the solution
$f_{\bm{k}=\bm{0}}(\bm{n})=\delta_{\bm{n},\bm{0}}$
contributes to the steady state. From Eqs.~\eqref{eq:wave_function}, \eqref{eq:wave_function2} and \eqref{eq:particle_number}, one finds $\lim_{t\to\infty}\ket{\psi(t)}=\frac{1}{L^d}\sum_{\bm{x}\in \Lambda} \hat{c}_{\bm{x}, \sigma}^{\dagger} \hat{c}_{\bm{x}, \tau}\ket{\mathrm{FM}_\tau}$ and $\lim_{t\to\infty} \mathcal{N}(t)=\frac{1}{L^d}$.

\bibliographystyle{apsrev4-1}
\bibliography{reference}

\end{document}